(21238) 1995 WV7: A New Basaltic Asteroid Outside the 3:1 Mean Motion Resonance

Mark Hammergren[1], Geza Gyuk[1,2], Andrew W. Puckett[2]

[1]Astronomy Department, Adler Planetarium & Astronomy Museum, 1300 S. Lake Shore Drive, Chicago, IL, 60605

[2]Department of Astronomy and Astrophysics, University of Chicago, 5640 S. Ellis Ave., Chicago, IL 60637

Manuscript text: 8 pages (double-spaced)

Figures: 2

Tables: 0

Abstract

We report visible to near-infrared spectroscopy and spectrophotometry of asteroid (21238) 1995 WV7 that reveal the presence of deep absorption bands indicating a V taxonomic type with an apparently basaltic surface composition. Since this asteroid is on the other side of the 3:1 mean motion resonance from Vesta, and because the required ejection velocity from Vesta is in excess of 1.6 km s$^{-1}$, we conclude that 21238 represents a sample of a differentiated body dynamically unrelated to Vesta.

Introduction

In 1970, spectral reflectance studies indicated the strong signature of the silicate mineral pyroxene on Vesta's surface, bearing striking similarities to spectra of the basaltic achondrite meteorites (McCord *et al.* 1970). Since then, further work has refined that basic picture, uncovering evidence for surface compositional heterogeneity correlating with topographic features, particularly an enormous crater near its south pole (Gaffey 1997, Cochran and Vilas 1998, Binzel *et al.* 1997, Thomas *et al.* 1997a, 1997b).

The large separation of Vesta from the strong 3:1 mean motion and $\nu_6$ secular resonance "meteorite supply routes", requiring ejection velocities of more than 600 m s$^{-1}$, was once viewed as an argument against Vesta as the parent body of the HED meteorites. Driven by the strong apparent compositional similarity between Vesta and the HED meteorites, however, some researchers simply used this separation as a constraint in their models of meteorite ejection from Vesta (Farinella *et al.* 1993). The discovery of small (~5-15 km) apparently basaltic asteroids bridging the gap between Vesta and the 3:1 mean motion resonance has been regarded by many as the "smoking gun" linking Vesta to the HED meteorites (Binzel and Xu 1993). Approximately 75 Vesta-like (or V-type) asteroids have been found in the main asteroid belt and among the near-Earth asteroid population (Tholen 1989, Xu *et al.* 1995, Lazzaro *et al.* 2000, Bus and Binzel 2002a, 2002b, Florczak *et al.* 2002, Binzel *et al.* 2004, Lazzaro *et al.* 2004, Marchi *et al.* 2005).

The discovery of a basaltic asteroid in the outer main belt, (1459) Magnya, that appears to be dynamically unrelated to Vesta (Lazzaro *et al.* 2000) has opened the door to studies of the remnants of other differentiated asteroids. In a detailed spectroscopic mineralogical analysis, Hardersen *et al.* (2004) find that Magnya is distinct from Vesta in orthopyroxene chemistry, concluding that the compositional difference precludes an origin on Vesta.

The discovery and detailed analysis of additional basaltic asteroids independent of Vesta would provide insights into the early history of solar system formation, particularly the pattern of heating and differentiation among the terrestrial planetary embryos. The identification of such objects is the major aim of our ongoing observational program.

Asteroids in the Sloan Digital Sky Survey

The Sloan Digital Sky Survey (SDSS) is a project to create a 10,000 deg$^2$ digital photometric and spectroscopic survey over the north galactic cap, along with a much deeper survey involving multiple scans in the southern galactic hemisphere (York *et al.* 2000). Although the primary focus of the SDSS is extragalactic studies, a large number of objects in the intervening space are also observed, including small bodies in the solar system.

Moving objects such as asteroids are detected by SDSS via their motion over the course of the five-filter exposure sequence, corresponding to an arc in time of about five minutes. While the time arc is too short to uniquely determine the orbits of previously unknown asteroids, it is possible to match known objects with SDSS moving object detections (Ivezi_ *et al.* 2001, Juri_ *et al.* 2002). The Third Release of the SDSS Moving Object Catalog, the latest publicly available version, contains data on 204,305 moving objects, including astrometric and photometric observations of 43,424 previously known objects (Ivezi_ *et al.* 2002).

Although the SDSS filters were not specifically chosen for asteroid reflectance studies, they have proven to be able to distinguish the major taxonomic types (Ivezi_ *et al.* 2001). In particular, asteroids exhibiting strong 0.9 _m absorption features, such as the V, A, and Q taxonomic types, have unusually blue (i'-z') colors relative to other asteroids.

Since 2005, we have conducted a program of visible-to-near-infrared spectroscopic confirmation of asteroids with unusually blue (i'-z') colors using the Dual Imaging Spectrograph on the ARC 3.5-m telescope at the Apache Point Observatory. Our target selection has been independent of that described in Roig and Gil-Hutton (2006), though naturally there is some overlap in target lists. As of August 2006, we have observed 58 asteroids. Those results are the subject of another paper (Hammergren *et al.*, in preparation). The majority of our targets have proven to be V-type objects. In this paper, we describe one of the results of our first half-night of observing, the confirmation of the V-type nature of (21238) 1995 WV7.

Observations and Results

We performed the observations on April 16, 2005 (UT) at the Apache Point Observatory, using the Dual Imaging Spectrograph on the Astrophysical Research Corporation 3.5-m telescope.

The DIS uses two cameras to simultaneously record the blue and red spectral regions. The low resolution blue grating and medium resolution red grating provided dispersions of 2.43 and 2.26 Å pixel$^{-1}$, respectively. The dichroic mirror has a transition at approximately 0.55 _m, causing strong variations in throughput extending to about 0.2 μm on either side. These variations are imperfectly removed in reduced spectra, so this immediate spectral region is excluded from our plots. The configuration permits the coverage of the entire spectral range from approximately 0.36 – 1.0 μm.

The 1.5-arcsecond wide spectrograph slit was maintained at the parallactic angle to minimize the effects of differential refraction. Solar analog stars were observed periodically to remove telluric absorptions and for production of the reflectance spectrum. We observed 21238 for a total of 4200 seconds, in three separate exposures, at between 1.22 and 1.36 airmasses.

Data reduction was performed using the Image Reduction and Analysis Facilty (IRAF) package, following standard procedures. After subtraction by an average bias frame and division by an average flat field, the spectra were optimally extracted using the *apextract* package. Wavelength calibration was performed using observations of helium, neon, and argon arcs. All spectra were corrected for atmospheric absorption using the average atmosphere for Kitt Peak, to minimize the effects of differences in airmass between the asteroid and solar analog observations. The spectrum of 21238 was divided by the spectrum of the solar analog star HD144873 (which was observed at an airmass of 1.2). Finally, the reflectance spectrum was normalized to unity at 0.55 μm by convention. No significant variations in the spectra between exposures were noted. The coadded, normalized reflectance spectrum of (21238) 1995 WV7 is shown in Fig. 1.

The SDSS photometry was converted to reflectance by reference to the solar colors of Ivezi ́ *et al.* (2001) and normalized such that the g band reflectance was in agreement with the reflectance spectrum.

Near infrared spectrophotometric observations were obtained on May 12, 2005 using the Near-Infrared Camera / Fabry Perot Spectrograph (NIC-FPS) on the ARC 3.5-m telescope at the Apache Point Observatory. Images were taken using broadband Mauna Kea J, H, and Ks filters, Gunn-Thuan z filter, and a narrowband [Si VI] filter centered at 1.965 _m (with a FWHM of 0.0065 _m). The observations were performed in the order z – J – H – Ks – [SiVI] – K – H – J – z to permit interpolation to minimize rotational lightcurve effects. The solar analog star P177D (Colina and Bohlin 1997) was also observed.

The instrumental magnitudes were corrected for atmospheric extinction using the following assumed primary extinction coefficients: $k_z$ = 0.06 (AJ 122, 2129); $k_J$ = 0.096, $k_H$ = 0.026, $k_K$ = 0.066 (2000 AJ 120, 3340); and an estimated $k_{[SIVI]}$ = 0.1 (a higher extinction than the K filter due to the [SiVI] filter's closer proximity to a water vapor absorption band). Because the difference in airmass between our observations of 21238 and P177D were never greater than about 0.22 (usually around 0.1), even if the assumed extinction coefficients were off by a factor of two, the systematic error in the reflectance spectrum would only amount to a few percent.

No significant difference was noted between early and late observations in each filter for 21238, so the lightcurve either has a long period or low amplitude. The magnitudes in each filter for 21238 and P177D were interpolated to the time of the midpoint of the respective object's [SiVI] sequence via linear fits in time. The solar analog star magnitudes were subtracted from the asteroid magnitudes (the equivalent of division in a linear flux scale to produce relative reflectances). Since we did not observe another standard star besides P177D, we report here only the magnitudes for 21238

relative to P177D: z = 3.82 ± 0.03; J = 3.34 ± 0.02; H = 3.69 ± 0.02; Ks = 3.70 ± 0.02; [SiVI] = 3.95 ± 0.11. These relative magnitudes were converted to linear reflectances, and normalized to the visible reflectance spectrum so that the z band reflectances were in agreement.

The combined reflectance spectrum of (21238) 1995 WV7 including both the visible-to-near-infrared spectrum and near-infrared spectrophotometry is displayed in Fig. 2. Also plotted is the spectrophotometric data for Vesta from Gaffey (1997). The presence of strong absorption features near 0.9 _m and 2.0 _m is characteristic of pyroxene and is a hallmark of the taxonomic V class. The dominance of pyroxene in V-type spectra is generally believed to indicate a basaltic surface composition and thus a high degree of igneous differentiation. A more quantitative analysis following the taxonomic scheme of Bus and Binzel (2002b) also returned a V-type classification.

Dynamics

Asteroid 21238 is located at a semimajor axis of 2.54 AU, just outside the 3:1 mean motion resonance, and on the opposite side of the resonance from Vesta ($a$ = 2.36 AU) and its associated family of V-type asteroids. This separation in semimajor axis alone would require an ejection velocity from Vesta of over 800 m s$^{-1}$ (Binzel and Xu 1993). Using a simplified inversion of Gauss's equations that takes into account the differences in semimajor axis, eccentricity, and inclination (Zappalá *et al.* 2002), we find that the minimum ejection velocity is greater than 1.6 km s$^{-1}$. These velocities are about an order of magnitude greater than the typical ejection velocities resulting from family-forming events, both as determined by studies of asteroid families (Nesvorn_ *et al.* 2006,

Vokrouhlick_ *et al.* 2006) and from the results of hydrocode simulations (Love and Ahrens 1996, Ryan and Melosh 1998, Benz and Asphaug 1999). Thus, we believe it is highly unlikely that 21238 has been placed into its current orbit directly from Vesta.

We also believe it is unlikely that 21238 migrated across the 3:1 resonance due to the Yarkovsky effect. With an absolute magnitude $H$ = 12.9, and conservatively assuming an albedo somewhere in the range of $p_v$ = 0.15 – 0.5, the diameter of 21238 is between 5 km – 9 km. At this size, the Yarkovsky drift rate in semimajor axis for a regolith-covered body is around $da/dt$ ~ 1 – 2 x $10^{-5}$ AU Myr$^{-1}$, assuming $da/dt \propto D^{-1}$ (Farinella *et al.* 1998). At the proper eccentricity of 21238 ($e$ = 0.1334), the 3:1 resonance is about 0.019 AU wide (Morbidelli and Vokrouhlick_ 2003). Assuming a uniform drift rate, this back-of-the-envelope analysis suggests that it would take 21238 more than one billion years to cross the resonance, which is more than two orders of magnitude greater than the dynamical lifetime of objects within the 3:1 resonance (Gladman *et al.* 1997).

We examined the possibility that 21238 was transported into its current location through the action of the 3:1 (or another) resonance, by integrating its orbit using the SWIFT software package (Levison and Duncan 1994). The integration included the effects of the Sun, the Earth/Moon system, and all other planets from Mercury through Pluto, as well as the four most massive asteroids, but neglected the Yarkovsky effect and other non-gravitational forces. We found that 21238 remained in a stable orbit and did not substantially migrate for two billion years, although there are a number of weak resonances in its neighborhood.

We also do not think it likely that 21238 evolved into its current orbit through the effects of close encounters with massive asteroids. First, in our integration, we saw no strongly-perturbing close encounters between 21238 and either (1) Ceres, (2) Pallas, (4) Vesta, or (10) Hygeia. Second, other researchers have shown that the perturbations due to close encounters with asteroids results in drifts much smaller than the width of the 3:1 resonance. In simulations spanning 100 Myr and including the 682 asteroids larger than 50 km in diameter, Carruba et al. (2003) find that the maximum displacement of any of their test particles in the intermediate main belt ($a$ = 2.6 – 2.8 AU, $e$ = 0.1, $i$ = 0 – 15°) was only $1.2 \times 10^{-3}$ AU, with most of these effects due to the four largest asteroids. Examination of their Figure 2 shows that at the location of 21238, the mean drift rate is less than $5 \times 10^{-6}$ AU Myr$^{-1}$, slower than the drift rate due to the Yarkovsky effect.

In the absence of an identified mechanism by which 21238 could have been moved into its current orbit from the neighborhood of Vesta, we conclude that 21238 likely has resided near its current location since its formation.

Conclusions

We present visible and near-infrared reflectance spectra and spectrophotometry that show that (21238) 1995 WV7 has a V-type taxonomy, with a surface apparently dominated by basalt. Dynamical studies indicate that 21238 was not ejected directly from Vesta, nor did it drift across the 3:1 mean motion resonance. We thus conclude that 21238 represents a fragment of a differentiated body independent of Vesta, making it only the second such object known besides (1459) Magnya.

Acknowledgements

Based on observations obtained with the Apache Point Observatory 3.5-meter telescope, which is owned and operated by the Astrophysical Research Consortium. This material is based upon work supported by the National Aeronautics and Space Administration under Grant No. NNG06GI40G issued through the Planetary Astronomy Program. M.H. and G.G. were supported in part by a grant from the Brinson Foundation.

Figure Captions:

Figure 1: Reflectance spectrum of (21238) 1995 WV7. A smoothed spectrum comprised of the median within 0.2 _m-wide bands is overplotted. The SDSS photometry, converted to reflectance values and normalized to coincide with the reflectance spectrum, is plotted as circles.

Figure 2: Reflectance spectrum of (21238) 1995 WV7, along with near-IR spectrophotometry. Spectrophotometry of 4 Vesta from Gaffey (1997) is plotted as gray circles for comparison.

Figure 1

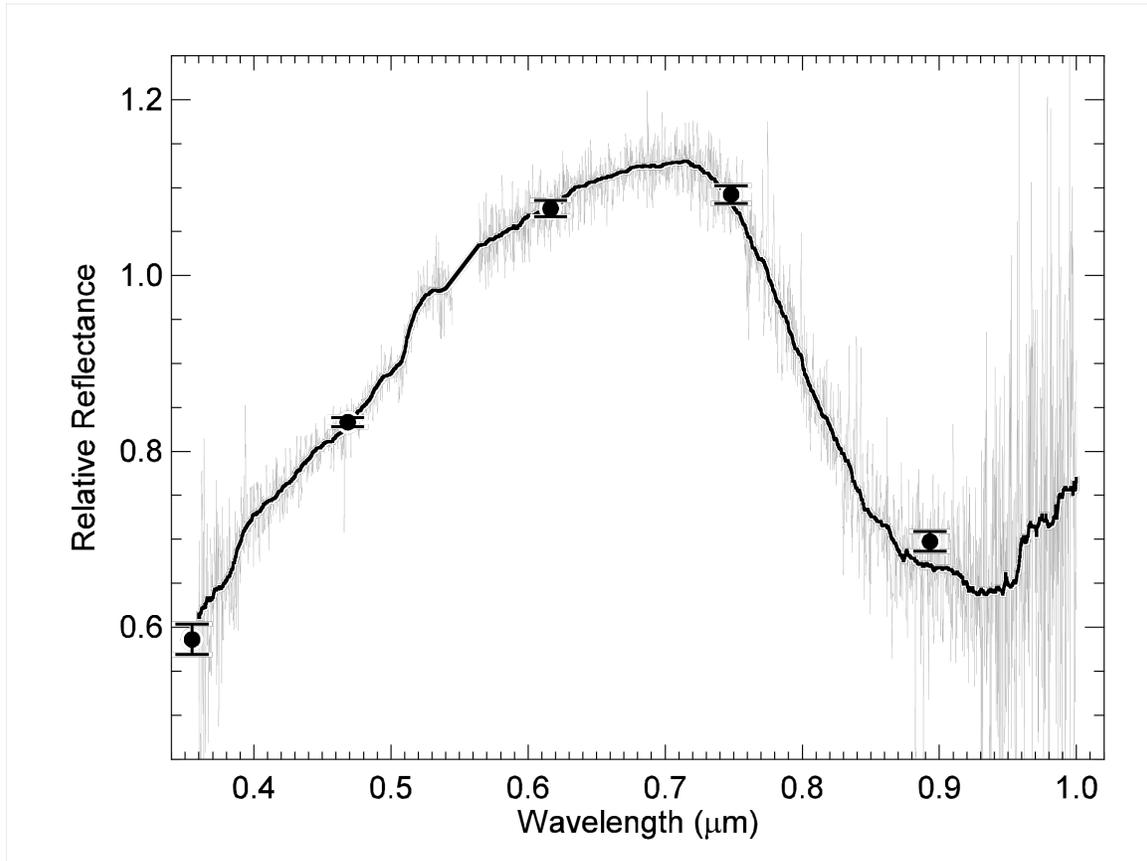

Figure 2

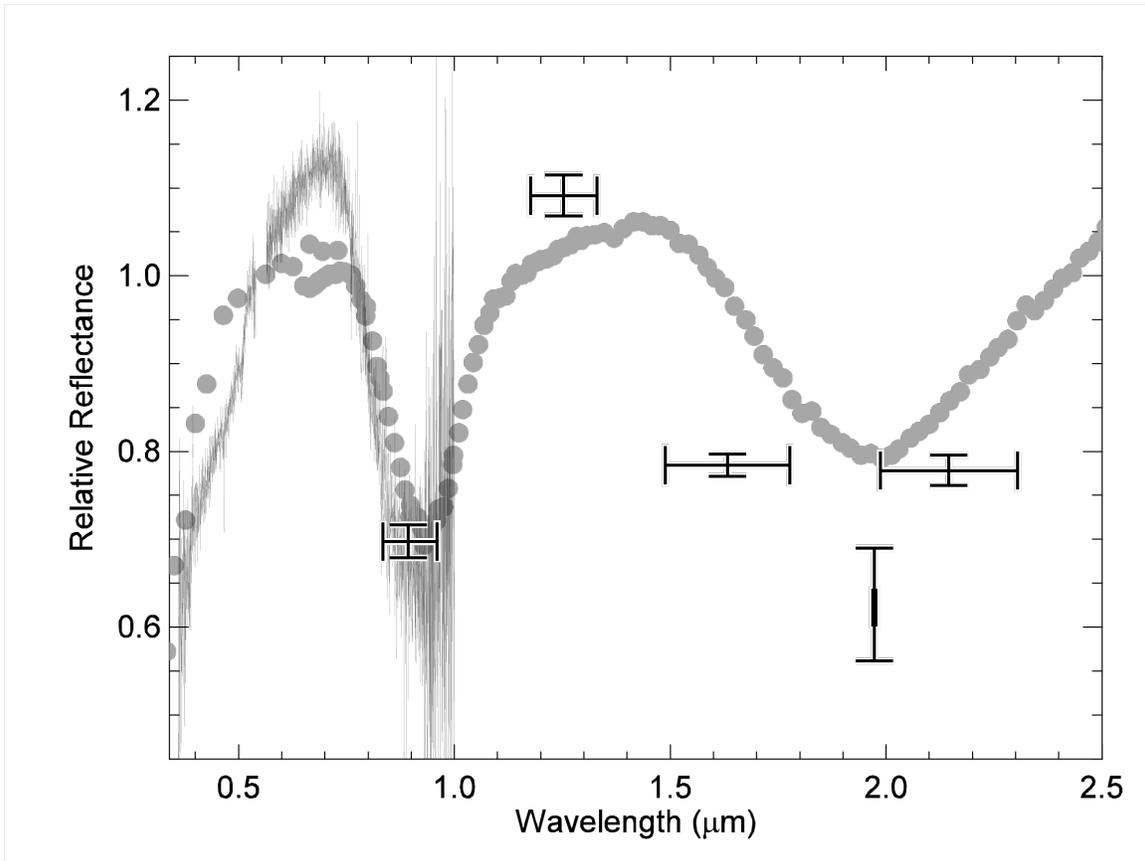